\documentclass[a4paper,twocolumn,english,aps,prl,showpacs,superscriptaddress]{revtex4}
\usepackage[T1]{fontenc}
\usepackage[latin1]{inputenc}
\usepackage{graphicx}
\usepackage{amssymb}

\makeatletter

\usepackage{epsfig}

\usepackage{babel}
\makeatother
\begin{document}

\title{Chaotic advection and targeted mixing}

\author{T. Benzekri%
\footnote{Permanent address: Universit\'{e} des Sciences et de Technologie
H.B., B.P. 32, El Alia, Bab Ezzouar, Alger, Algeria%
}}

\affiliation{Centre de Physique Th\'{e}orique%
\footnote{Unit\'{e} Mixte de Recherche (UMR 6207) du CNRS, et des universit\'{e}s
Aix-Marseille I, Aix-Marseille II et du Sud Toulon-Var. Laboratoire
affili\'{e} \`{a} la FRUMAM (FR 2291).%
}, CNRS Luminy, Case 907, F-13288 Marseille cedex 9, France}

\author{C. Chandre}

\affiliation{Centre de Physique Th\'{e}orique%
\footnote{Unit\'{e} Mixte de Recherche (UMR 6207) du CNRS, et des universit\'{e}s
Aix-Marseille I, Aix-Marseille II et du Sud Toulon-Var. Laboratoire
affili\'{e} \`{a} la FRUMAM (FR 2291).%
}, CNRS Luminy, Case 907, F-13288 Marseille cedex 9, France}

\author{X. Leoncini}

\affiliation{PIIM, Universit\'{e} de Provence, Case 321, Centre de Saint J\'{e}r\^{o}me,
F-13397 Marseille cedex 20, France}

\author{R. Lima}

\affiliation{Centre de Physique Th\'{e}orique%
\footnote{Unit\'{e} Mixte de Recherche (UMR 6207) du CNRS, et des universit\'{e}s
Aix-Marseille I, Aix-Marseille II et du Sud Toulon-Var. Laboratoire
affili\'{e} \`{a} la FRUMAM (FR 2291).%
}, CNRS Luminy, Case 907, F-13288 Marseille cedex 9, France}

\author{M. Vittot}

\affiliation{Centre de Physique Th\'{e}orique%
\footnote{Unit\'{e} Mixte de Recherche (UMR 6207) du CNRS, et des universit\'{e}s
Aix-Marseille I, Aix-Marseille II et du Sud Toulon-Var. Laboratoire
affili\'{e} \`{a} la FRUMAM (FR 2291).%
}, CNRS Luminy, Case 907, F-13288 Marseille cedex 9, France}

\begin{abstract}
The advection of passive tracers in an oscillating vortex chain is
investigated. It is shown that by adding a suitable perturbation to
the ideal flow, the induced chaotic advection exhibits two remarkable
properties compared with a generic perturbation~: Particles remain
trapped within a specific domain bounded by two oscillating barriers
(suppression of chaotic transport along the channel), and the stochastic
sea seems to cover the whole domain (enhancement of mixing within
the rolls). 
\end{abstract}

\date{\today{}}


\maketitle
In order to enhance mixing in flows one often relies on chaotic advection
\cite{Ar}. This phenomenon translates the fact that despite the laminar
character of a given flow, trajectories of fluid particles or advected
passive tracers are chaotic. As a consequence, mixing is considerably
enhanced in regions where chaos is at play, as it does not need to
rely on molecular diffusion. Such phenomena are observed in a wide
range of physical systems \cite{Ot} and have fundamental applications,
for instance, in geophysical flows \cite{BeMeSw,BrSm} or chemical
engineering. When considering two dimensional incompressible flows,
chaotic advection occurs when the flow is unsteady, see for instance
Refs.~\cite{SoGo,SoMiSpMo,WiCaTA,CaWi}. One peculiarity of these
flows is that the dynamics of passive particles can be tackled from
the Hamiltonian dynamics point of view. The canonical variables are
directly the space variables, hence the phase space is formally the
two dimensional physical space. This feature allows a direct visualization
of the phase space in experiments, and in a certain sense makes this
framework an ideal one in order to confront theoretical results on
Hamiltonian dynamics with experiments. 

In this Letter, we address the problem of targeted mixing. Namely
it is often desirable to reduce chaotic transport (see Ref.~\cite{CCDLMV})
while one also often wants more chaos in order to enhance mixing (see,
for instance, Refs.~\cite{Ba,Mez} for microfluidic and microchannel
devices). In order to achieve such properties, we consider the dynamics
of passive tracers in an array of alternating vortices. This flow
has been realized experimentally using Rayleigh-B\'{e}nard convection
or in a more controlled way by using electromagnetic forces~\cite{SoGo,SoMiSpMo,WiCaTA}.
The primary interest in this flow resides in the fact that being generated
by quite a few hydrodynamic instabilities, it may be considered as
one of the founding bricks of turbulence. As such, understanding its
influence on the advection of passive or active quantities is considered
a necessary first step in order to uncover the different mechanisms
governing for instance front propagation in turbulent flows~\cite{AbCeVeVu,PoHa}. 

\begin{figure}
\includegraphics[%
  width=5cm]{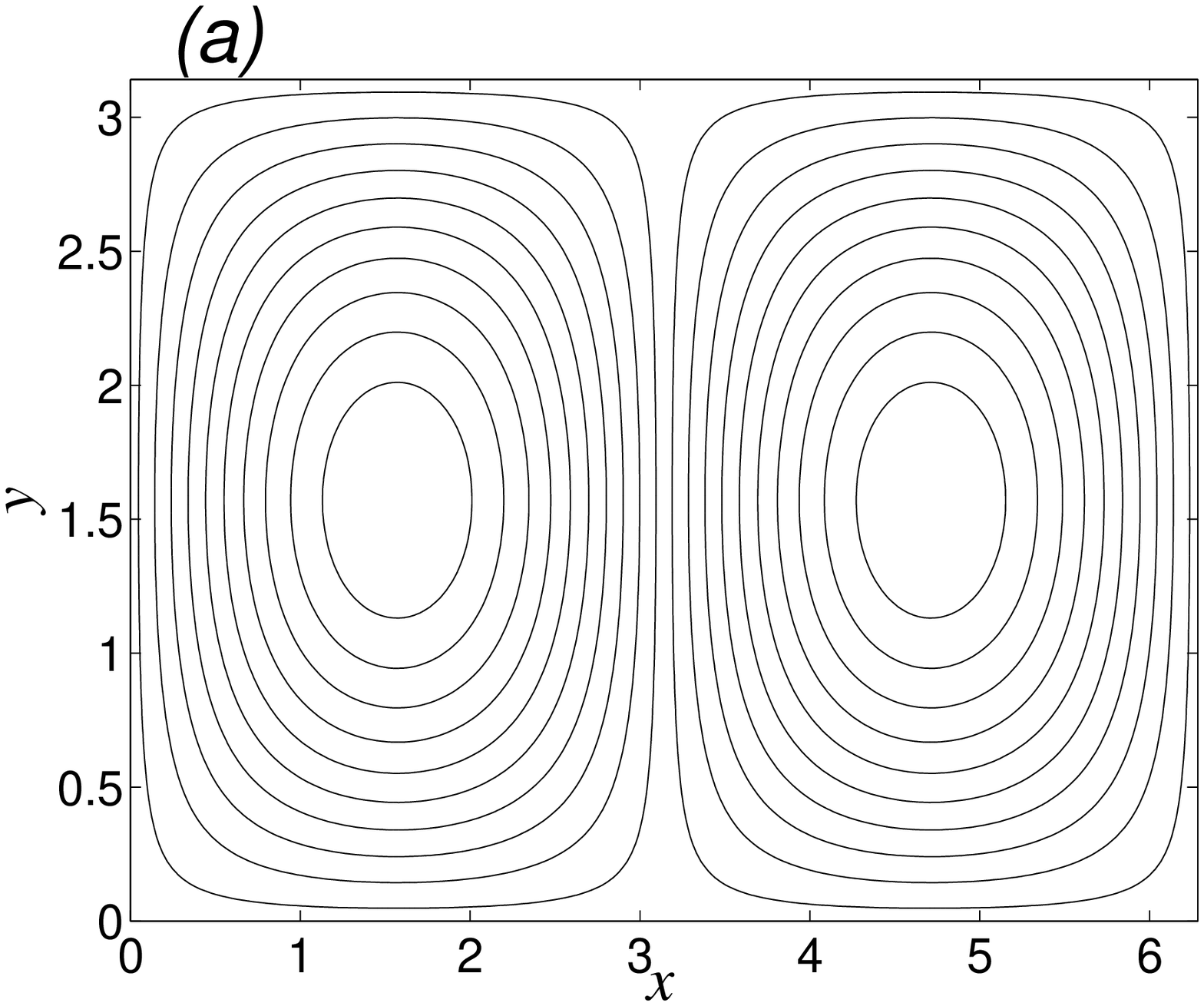}

\includegraphics[%
  width=5cm]{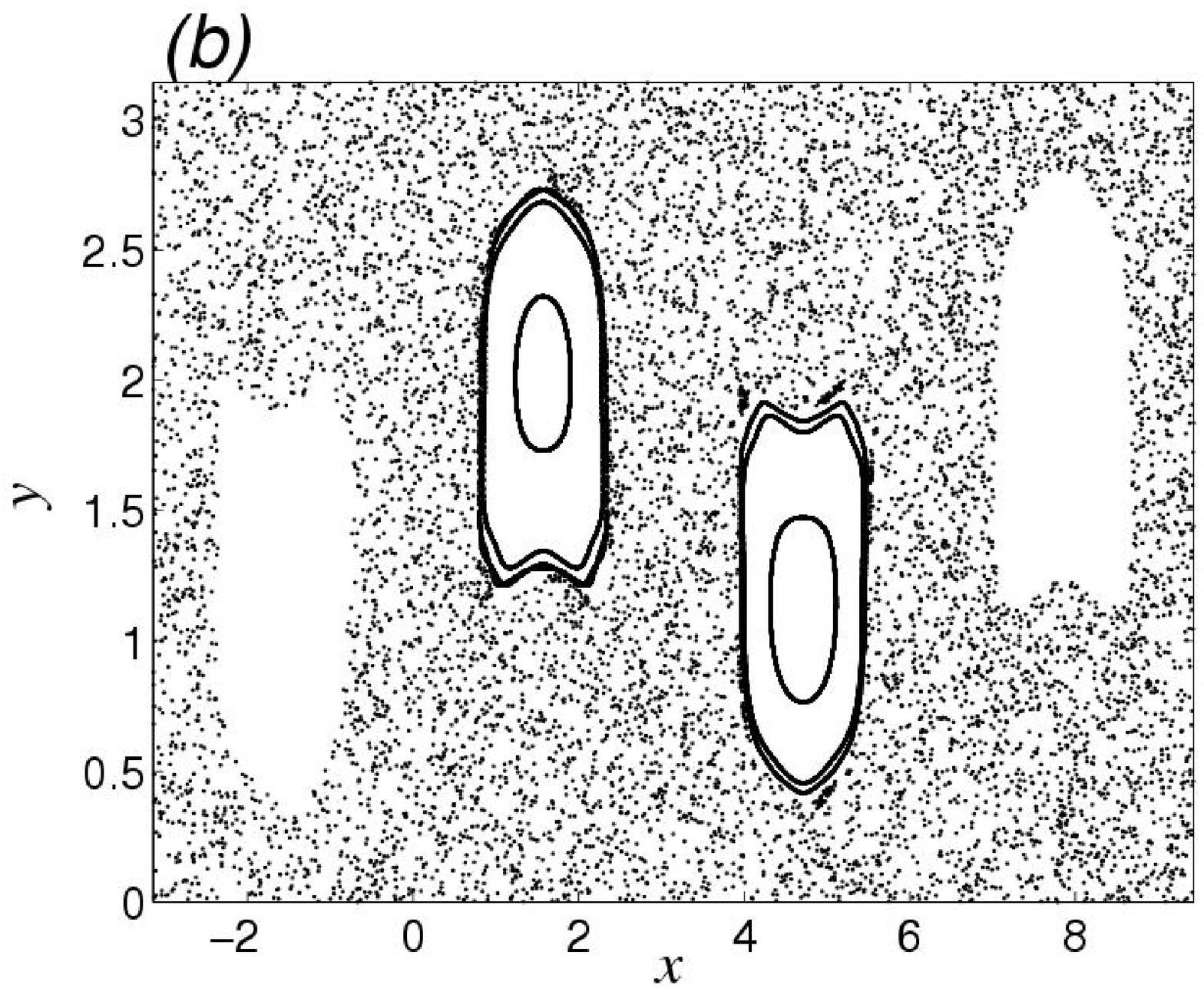}

\caption{\label{Fig1} Poincar\'{e} sections $(a)$ for the stream function~(\ref{psi0})
which are also the streamlines and $(b)$ for the stream function~(\ref{courant}).
The parameters are $\alpha=0.6$ and $\epsilon=0.63$.}
\end{figure}

We consider the following integrable stream function which models
the experiment with slip boundary conditions~: \begin{equation}
\Psi_{0}(x,y)=\alpha\sin x\sin y,\label{psi0}\end{equation}
 where the $x$-direction is the horizontal one along the channel
and the $y$-direction is the bounded vertical one. The constant $\alpha$
is the maximal value of vertical velocity. Passive particles follow
the streamlines of $\Psi_{0}$ depicted on Fig.~\ref{Fig1}$(a)$.
The dynamics given by Eq.~(\ref{psi0}) is integrable. Therefore
no mixing occurs. The fluid is limited by two invariant surfaces $y=\pi$
and $y=0$ corresponding to the top and bottom roll boundaries. The
flow has hyperbolic fixed points on these two invariant surfaces which
are localized at $x=m\pi$ for $m\in\mathbb{Z}$. These points are
joined by vertical heteroclinic connections for which the stable and
unstable manifolds coincide. The phase space, which is here the real
space, is then characterized by a chain of rolls with separatrices
localized at $x=m\pi$ for $m\in\mathbb{Z}$. 

In the experiments~\cite{SoGo,SoMiSpMo,WiCaTA}, a typical perturbation $f(x,y,t)$ is introduced
as a time dependent forcing in order to trigger chaotic advection
and then to study the resulting transport and mixing properties. More
precisely the perturbation modifies the stream function as~: \begin{equation}
\Psi_{c}(x,y,t)=\Psi_{0}(x+f(x,y,t),y).\label{psic}\end{equation}
 For instance, the following stream function has been proposed to
model an experimental situation~\cite{SoGo}~: \begin{equation}
\Psi_{1}(x,y,t)=\alpha\sin(x+\epsilon\sin\omega_{0}t)\sin y,\label{courant}\end{equation}
 i.e.\ the perturbation is $f=\epsilon\sin\omega_{0}t$ and describes
the lateral oscillations of the roll patterns where $\epsilon$ and
$\omega_{0}$ are respectively the amplitude and the angular frequency
of the lateral oscillations. Without loss of generality, we assume
that $\omega_{0}=1$. The resulting streamlines which are again closed
curves correspond to lateral oscillations in the $x$-direction of
the streamlines depicted in Fig.~\ref{Fig1}$(a)$ with a periodic
displacement of $-\epsilon\sin t$. However chaotic advection is triggered.

In order to infer transport and mixing properties we use Poincar\'{e}
sections of some trajectories of tracers. A typical Poincar\'{e}
section associated with the stream function $\Psi_{1}$ given by Eq.~(\ref{courant})
is represented in Fig.~\ref{Fig1}$(b)$ for $\epsilon=0.63$ and
$\alpha=0.6$. It shows that the transport along the channel is greatly
enhanced~\cite{SoGo} compared with no transport in Fig.~\ref{Fig1}$(a)$. The
advected tracers have now a non integrable Hamiltonian dynamics. The reason being that, under
the perturbation, the vertical heteroclinic connections between the
vortices break down and the stable and unstable manifolds intersect
transversely thus generating chaotic advection of passive tracers
along the channel~\cite{Wi}. Therefore the mixing properties of
the flow are enhanced. However, most of the time islands of stability
remain, among which the ones located around the centers of the vortices.
They are characterized by regular (quasiperiodic) trajectories and
a chaotic region around and between the rolls. Mixing within the center
of the rolls can only rely on molecular diffusion while tracers in
the chaotic sea are advected unboundedly along the channel and chaotic
mixing occurs. The main observation is that the regular patterns persist
and chaos or mixing are not well developed. 

In order to obtain what we called targeted mixing, we propose to adjust
the time-dependent forcing $f$ which we assume to depend on $y$
and $t$ such that the stream function $\Psi_{c}$ given by Eq.~(\ref{psic})
has the following properties~: First its attendant chaotic transport
along the channel ($x$-direction) is confined by impenetrable barriers.
Second, when restricted to each of these independent cells, we want to obtain good mixing properties. 

A first possibility to create barriers given by invariant tori of
the dynamics would be to use the method proposed in Refs.~\cite{ViChCiLi,CCDLMV}
for controlling Hamiltonian systems. This method uses an affine control
(modifying the Hamiltonian or the stream function additively). However
applying this method to the system under consideration would create
some barriers but would also regularize the inside of the cell, thus
drastically reducing the mixing. 

In order to proceed in finding a more suitable perturbation $f(y,t)$,
we start from a more generic situation for which the stream function
given by Eq.~(\ref{psi0}) is a particular case. We consider a generic
integrable stream function $\Psi_{0}(x,y)$ which is $L$-periodic
in $x$, and such that $\partial_{x}\Psi_{0}(x,0)=\partial_{x}\Psi_{0}(x,H)=0$
for all $x\in{\mathbb{R}}$ where $H$ is the height of the channel.
The perturbed stream function is written as in Eq.~(\ref{psic})
where $f(y,t)$ is such that $(i)$ there are barriers at $x=x_{k}(y,t)$
(for $k\in{\mathbb{Z}}$) which prevent the chaotic advection along
the $x$-direction and $(ii)$ there is a high mixing inside a region
bounded by two consecutive barriers, e.g., $x_{1}$ and $x_{2}$.
We notice that the perturbed stream function is obtained by adding
the perturbation to the canonical variables of the stream function
and not to the stream function itself as in Refs.~\cite{ViChCiLi,CCDLMV}.
The equations of the barriers are written as $x=x_{k}(y,t)=kL+\varphi(y,t)$
for $k\in{\mathbb{Z}}$ where $\varphi$ is a function to be determined.
Using Hamilton's equations \begin{eqnarray*}
\dot{x} & = & -\frac{\partial\Psi_{c}}{\partial y}=-\frac{\partial\Psi_{0}}{\partial y}(x+f,y)-\frac{\partial f}{\partial y}\frac{\partial\Psi_{0}}{\partial x}(x+f,y)\:,\\
\dot{y} & = & \frac{\partial\Psi_{0}}{\partial x}(x+f,y)\:,\end{eqnarray*}
 and by imposing that $x_{k}$ locates a barrier we get for the particles
on the barrier\[
\dot{x}=\frac{\partial\varphi}{\partial t}+\dot{y}\frac{\partial\varphi}{\partial y}\:.\]
 Thus a possible solution is obtained when $f+\varphi$ is only a
function of time, denoted $\Phi(t)$, and the condition on $\varphi$
is \[
\frac{\partial\varphi}{\partial t}=-\frac{\partial\Psi_{0}}{\partial y}(\Phi(t),y).\]
 The above equation is solved as \[
\varphi(y,t)=-\Gamma\partial_{y}\Psi_{0}(\Phi(t),y),\]
 where the linear operator $\Gamma$ is a pseudo-inverse of $\partial_{t}$,
i.e. acting on $v(y,t)=\sum_{k}v_{k}{\mathrm{e}}^{ikt}$ as \[
\Gamma v=\sum_{k\not=0}\frac{v_{k}}{ik}{\mathrm{e}}^{ikt}.\]
 The perturbation $f$ is then \begin{equation}
f(y,t)=\Phi(t)+\Gamma\partial_{y}\Psi_{0}(\Phi(t),y),\label{eqn:fct}\end{equation}
 where $\Phi$ is any function of time. 

We now apply these results to the array of vortices given by the stream
function (\ref{psi0}). In order to remain as close as possible to
the experimental setup, we consider $\Phi(t)=\epsilon\sin t$. %
\begin{figure}
\includegraphics[%
  width=5cm]{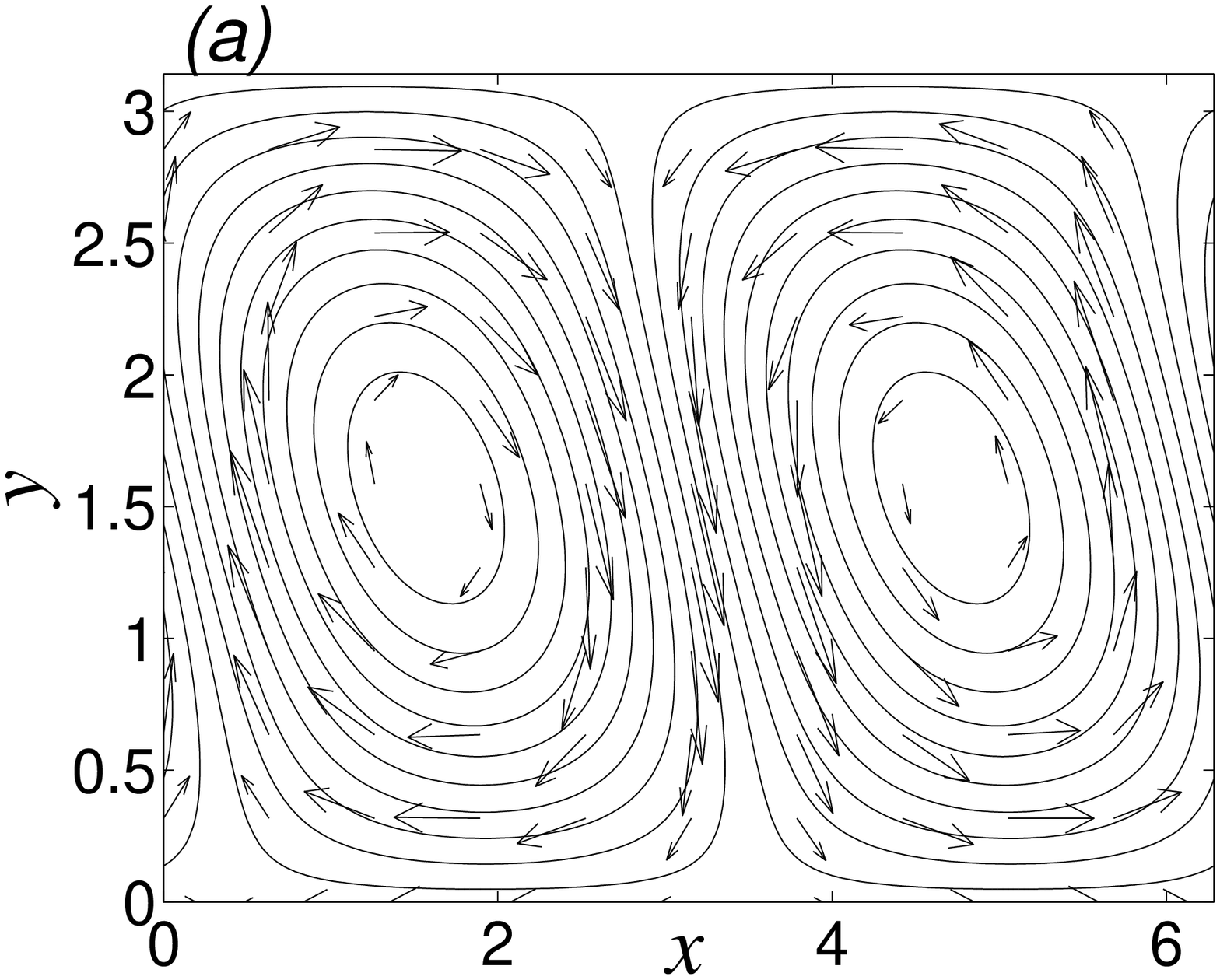}

\includegraphics[%
  width=5cm]{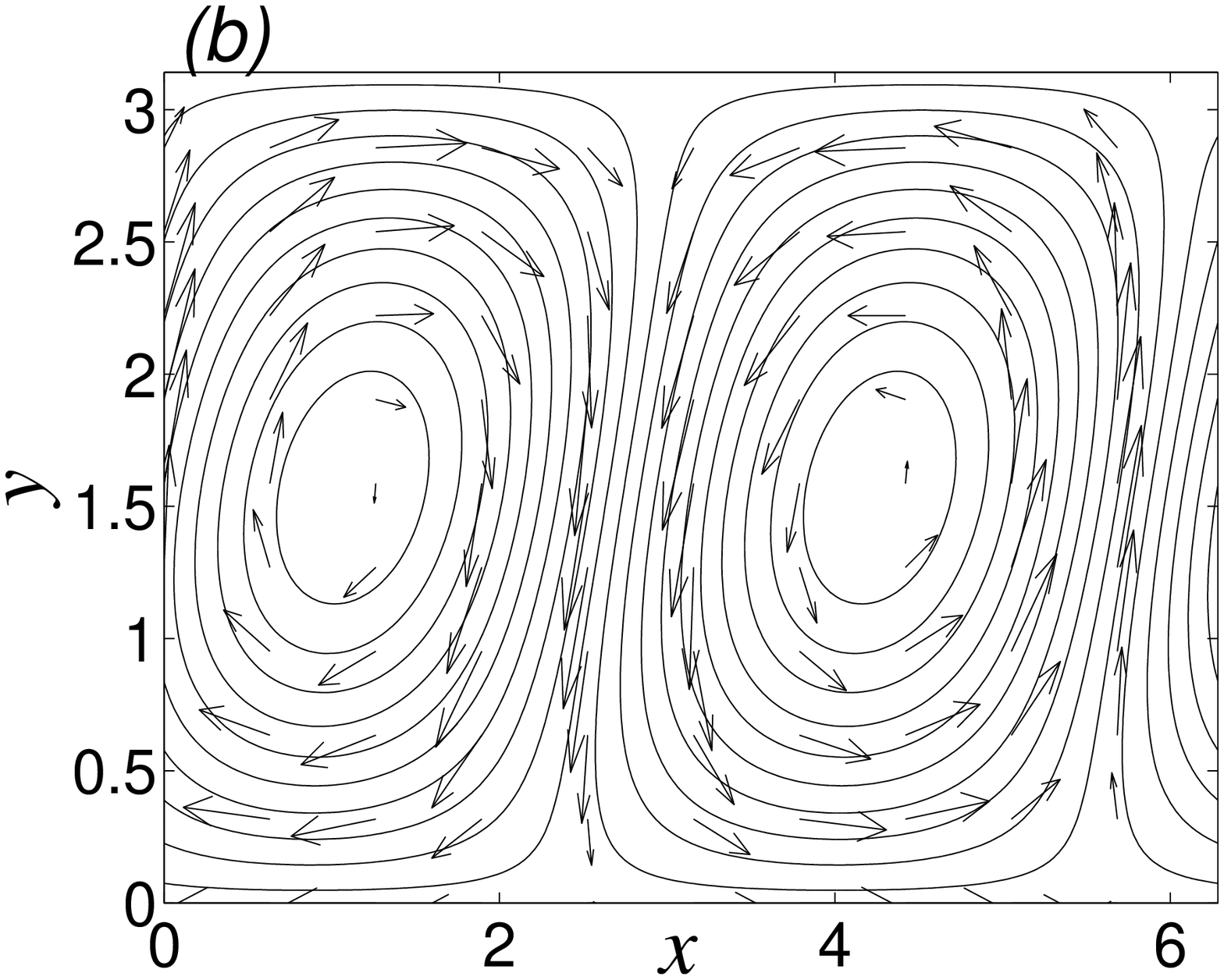}

\caption{\label{Fig2} Streamlines of the stream function~(\ref{Strmcont})
at $(a)$ $t=0$ and $(b)$ $t=3\pi/4$. The parameters are $\alpha=0.6$
and $\epsilon=0.63$.}
\end{figure}
The modified stream function given by Eq.~(\ref{eqn:fct}) is then
\begin{equation}
\Psi_{c}(x,y,t)=\alpha\sin[x+\epsilon\sin t+\alpha\cos yC_{\epsilon}(t)]\sin y,\label{Strmcont}\end{equation}
 where \begin{equation}
C_{\epsilon}(t)=\sum_{n\geq0}\frac{-2}{2n+1}\mathcal{J}_{2n+1}(\epsilon)\cos(2n+1)t,\label{serbessl}\end{equation}
 and $\mathcal{J}_{l}$ (for $l\in\mathbb{N}$) are Bessel functions
of the first kind. We notice that $y=0$ and $y=\pi$ are still boundaries
for the streamfunction ~(\ref{Strmcont}) which comes from the fact
that the modification is only on the $x$-term of the original stream
function. The streamlines are then slightly modified (non-uniformly
in $y$) as it can be seen on Fig.~\ref{Fig2}$(a)$ and $(b)$ which
depict the streamlines of the stream function~(\ref{Strmcont}) for
$\epsilon=0.63$ and $\alpha=0.6$ at two different times $t=0$ and
$t=3\pi/4$ respectively. Moreover, the displacement of these rolls
remains parallel to the $x$-axis as it is for the stream function
$\Psi_{1}$ given by Eq.~(\ref{courant}). The fact that the streamlines
of the stream function~(\ref{courant}) and the ones of Eq.~(\ref{Strmcont})
look similar comes from the fact that the stream function $\Psi_{c}$
is a small modification of $\Psi_{1}$ (for $\alpha$ small) since
we have $\left|\Psi_{c}(x,y,t)-\Psi_{1}(x,y,t)\right|\leq{\alpha^{2}}\epsilon/{2}$. 

Nevertheless, the dynamics of tracers is completely different in both cases
and targeted mixing is achieved. For the same values of $\epsilon$
and $\alpha$ as in Fig.~\ref{Fig1}$(b)$,%
\begin{figure}
\includegraphics[%
  width=7.5cm]{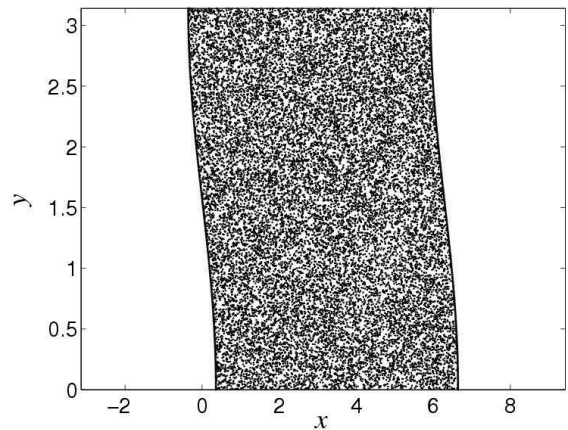}

\caption{\label{Fig3} Poincar\'{e} section for the stream function~(\ref{Strmcont}).
The parameters are $\alpha=0.6$ and $\epsilon=0.63$.}
\end{figure}
 a Poincar\'{e} section of the dynamics of passive tracers in a flow
described by the stream function $\Psi_{c}$ given by Eq.~(\ref{Strmcont})
is represented in Fig.~\ref{Fig3}. We notice that there are invariant
surfaces which have been created around $x=0$ (mod $2\pi$) (bold
curves) and that mixing within these barriers is considerably enhanced.
The equations of these transport barriers along the $x$-direction
are~: \begin{equation}
x=x_{k}(y,t)=2k\pi-\alpha\cos yC_{\epsilon}(t).\label{eqbar}\end{equation}
 To be more specific, the barriers~(\ref{eqbar}) are degenerate
invariant tori since each of them is a heteroclinic connexion between
two hyperbolic periodic orbits (with period $2\pi$), e.g., one locate
at $y=0$ and $x(t)=-\alpha C_{\epsilon}(t)$ and the other one at
$y=\pi$ and $x(t)=\alpha C_{\epsilon}(t)$, moving in opposite directions
along the channel. Regarding the enhancement of mixing inside the
cell bounded by the two barriers observed in Fig.~\ref{Fig3}, we
notice that most of the regular trajectories observed for the stream
function $\Psi_{1}$ are broken by the perturbation (in comparison
with Fig.~\ref{Fig1}$(b)$). It means that, the perturbation given
by Eq.~(\ref{eqn:fct}) creates two invariant surfaces around $x=0$
and $x=2\pi$, and destabilizes the inside of the cell and in particular
the remainders of the regular motion near $x=\pi/2$ and $x=3\pi/2$. 

In order to get some clues on the origin of mixing enhancement, we
consider the region in the middle of the cell of length $L$ in the
$x$-direction and compute the Melnikov function which measures the
distance between the stable and unstable manifolds. First, we perform
the (time-dependent) canonical transformation $(x,y)\mapsto(x',y')$
generated by $F_{2}(y,x',t)=yx'-\Gamma\Psi_{0}(\Phi(t),y)$. In the
case of $\Psi_{0}$ given by Eq.~(\ref{psi0}), the stream function
becomes \[
\tilde{\Psi}_{c}(x',y',t)=\alpha[\sin(x'+\Phi(t))-\sin(\Phi(t))]\sin y'.\]
 The Melnikov function is defined as~\cite{CaWi} \[
M(t_{0})=\int_{-\infty}^{+\infty}\{\Psi_{0},\tilde{\Psi}_{c}\}(x_{0}(t-t_{0}),y_{0}(t-t_{0}))\, dt,\]
 where $(x_{0}(t),y_{0}(t))$ is a solution of the unperturbed system
$\Psi_{0}$. The Poisson bracket between $\Psi_{0}$ and $\tilde{\Psi}_{c}$
is given by \[
\{\Psi_{0},\tilde{\Psi}_{c}\}=\alpha^{2}(1-\cos x)\sin(\Phi(t))\sin y\cos y.\]
 It is straightforward to see that the Melnikov function vanishes
when $x=0\mbox{ mod }2\pi$, i.e.\ on the boundaries of the cell,
and is maximum (in absolute value) when $x=\pi\mbox{ mod }2\pi$,
i.e.\ in the middle of the cell. Therefore it is expected that the mixing is enhanced since then, there is a maximum flux inside the cell~\cite{Ba}. 

In order to test the robustness of the method and to try an experimentally
more tractable perturbation, we truncate the series given by Eq.~(\ref{serbessl}). The simplest possible perturbation is obtained by considering only the first term of the series
$C_{\epsilon}(t)$ in Eq.~(\ref{serbessl}) which leads to the following
stream function \begin{equation}
\Psi_{s}(x,y,t)=\alpha\sin\left(x+\epsilon\sin t-2\alpha\mathcal{J}_{1}(\epsilon)\cos y\cos t\right)\sin y,\label{psis}\end{equation}
 which has a much simpler time-dependence. We performed a Poincar\'{e} section for
this simplified stream function $\Psi_{s}$ given by Eq.~(\ref{psis})
for $\alpha=0.6$ and $\epsilon=0.63$. This Poincar\'{e} section
looks identical to the one depicted on~\ref{Fig3} showing that the
simplified perturbation remains efficient. %
\begin{figure}
\includegraphics[%
  width=3cm]{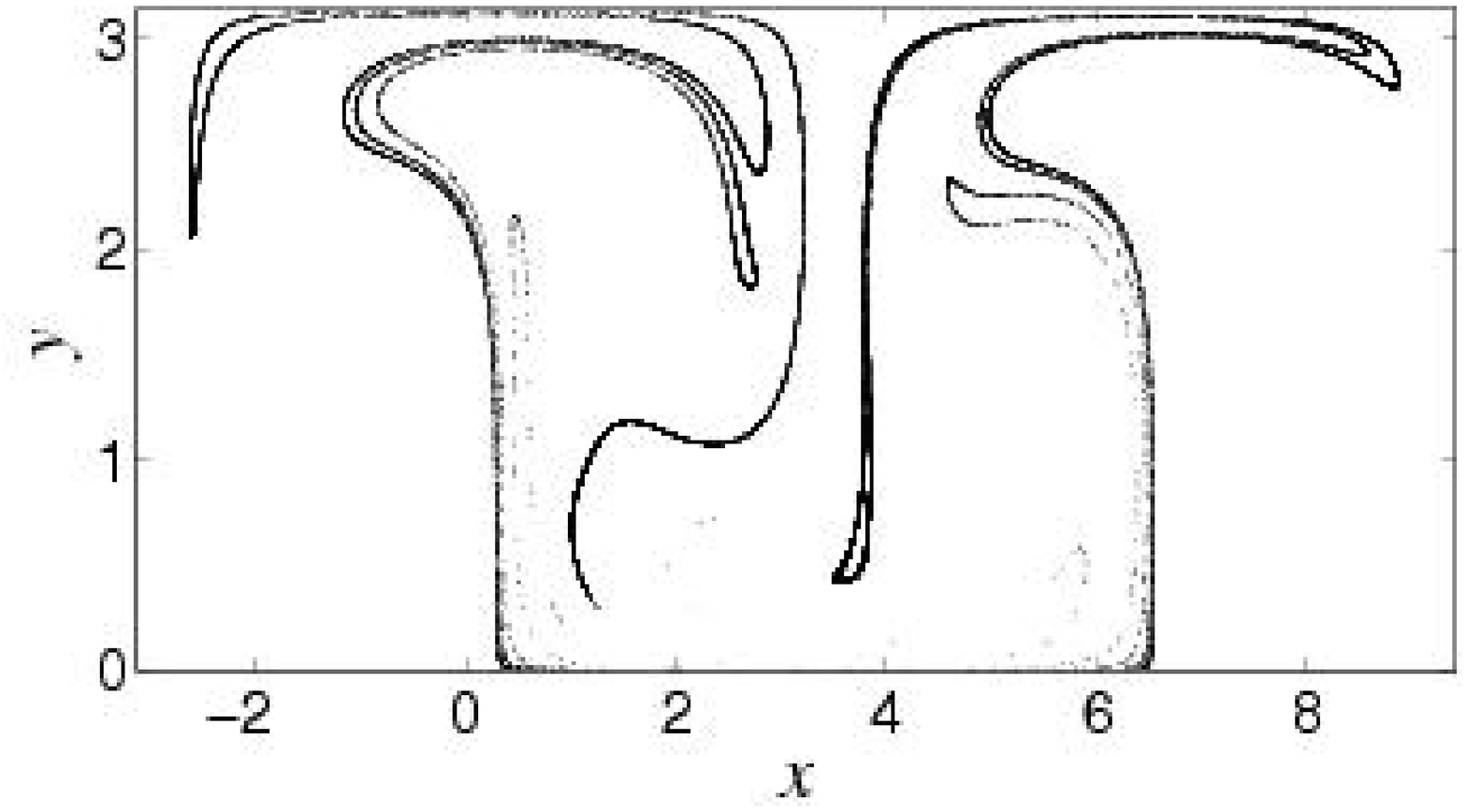}\includegraphics[%
  width=3cm]{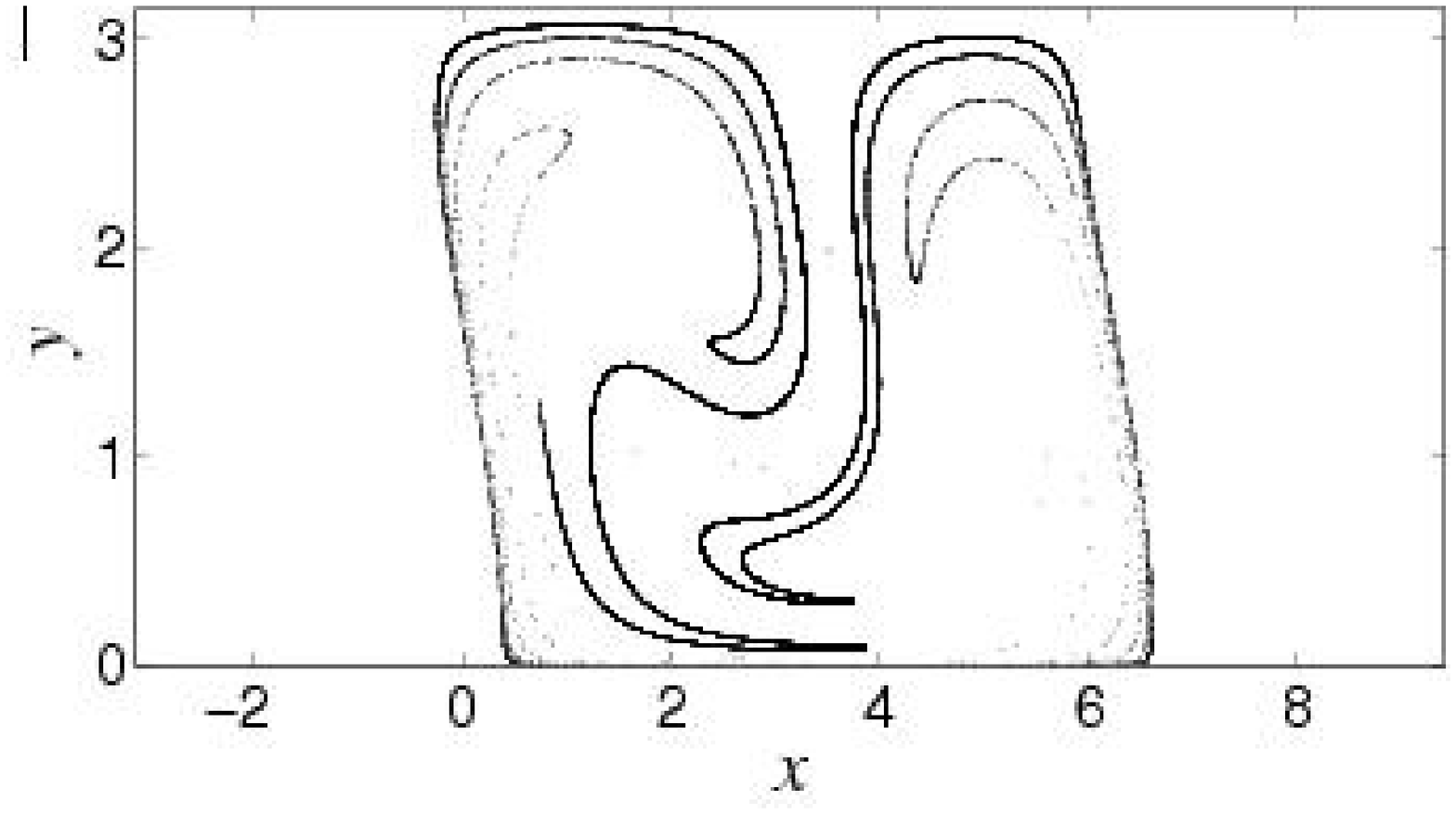}

\includegraphics[%
  width=3cm]{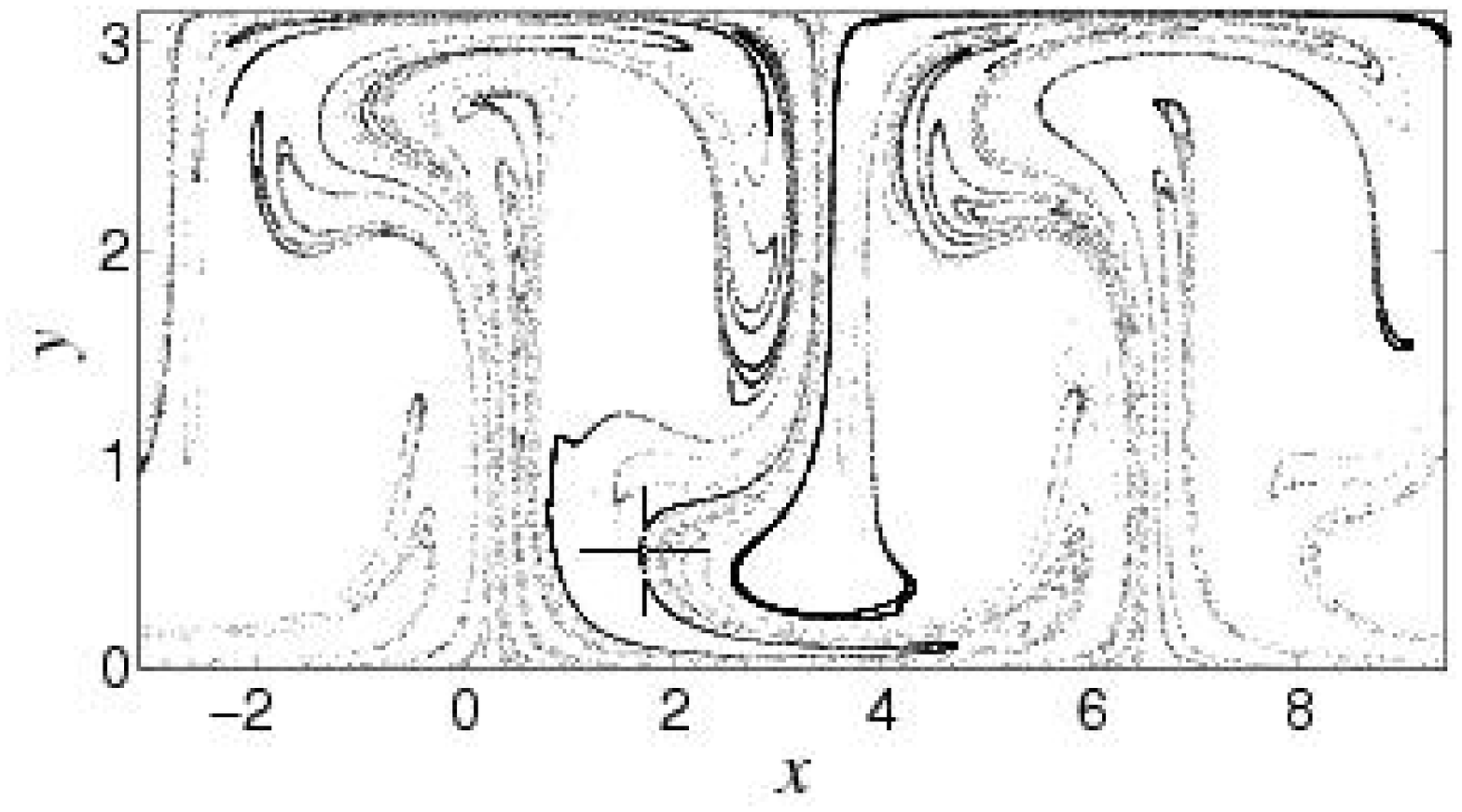}\includegraphics[%
  width=3cm]{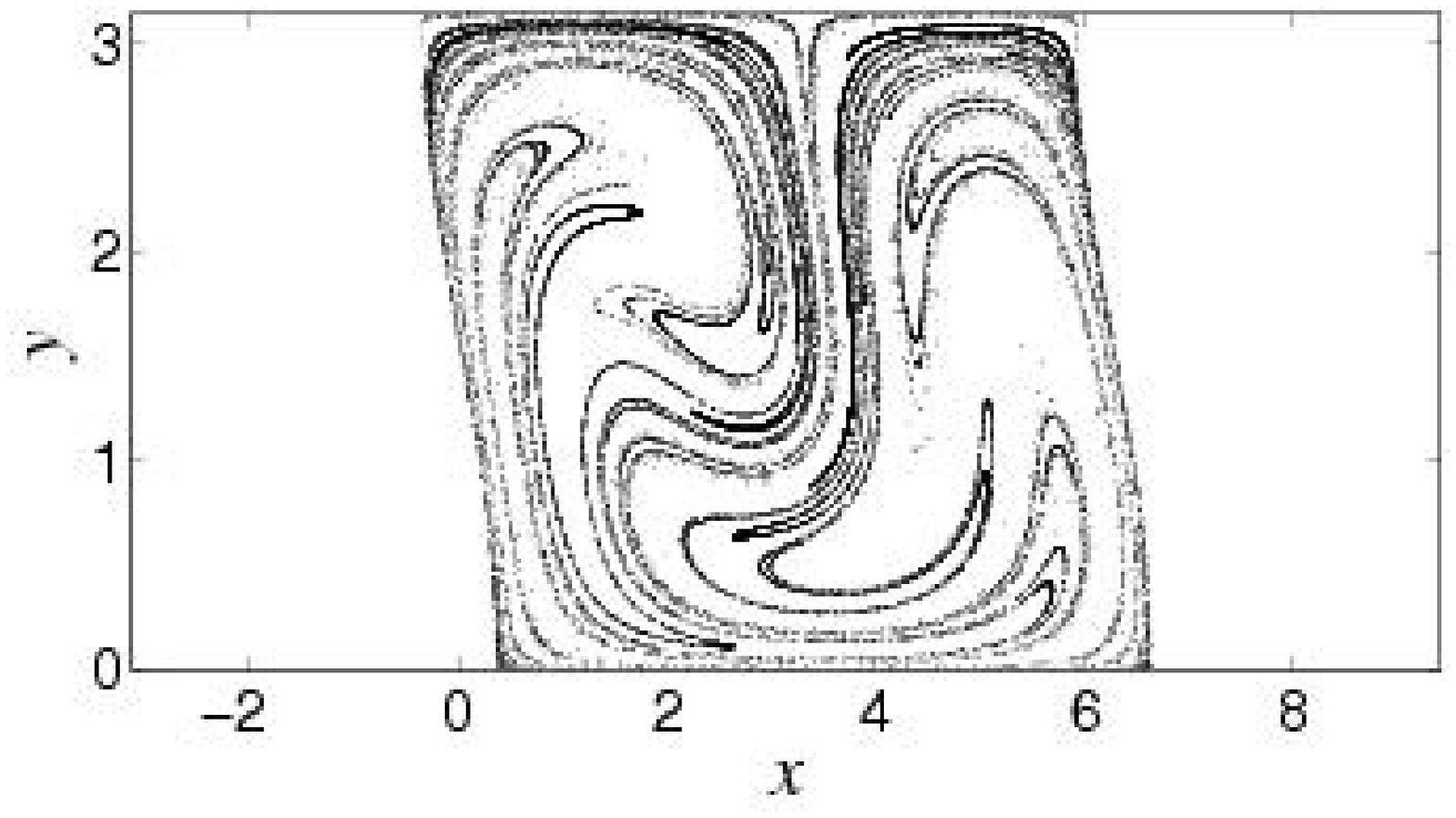}

\includegraphics[%
  width=3cm]{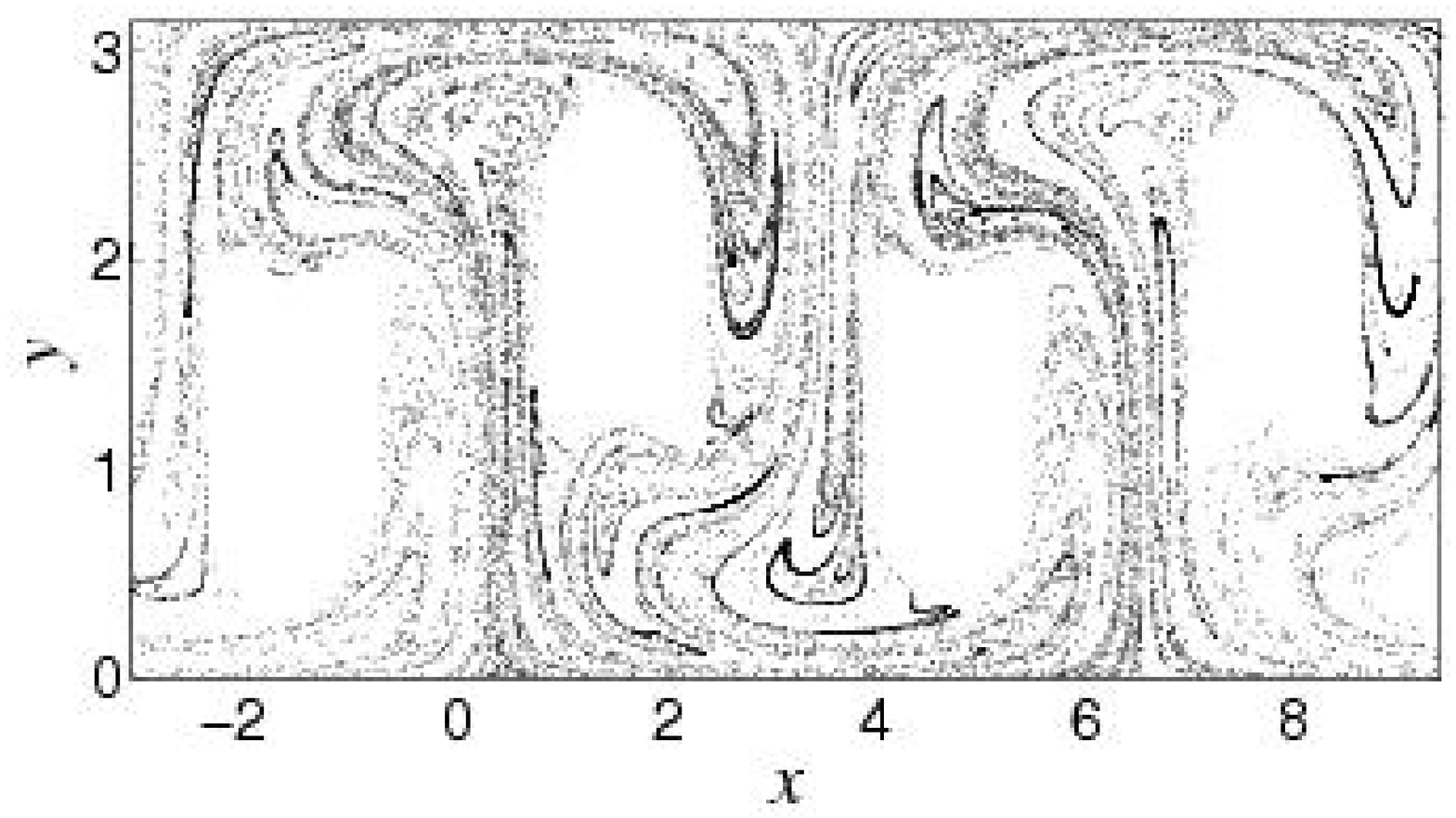}\includegraphics[%
  width=3cm]{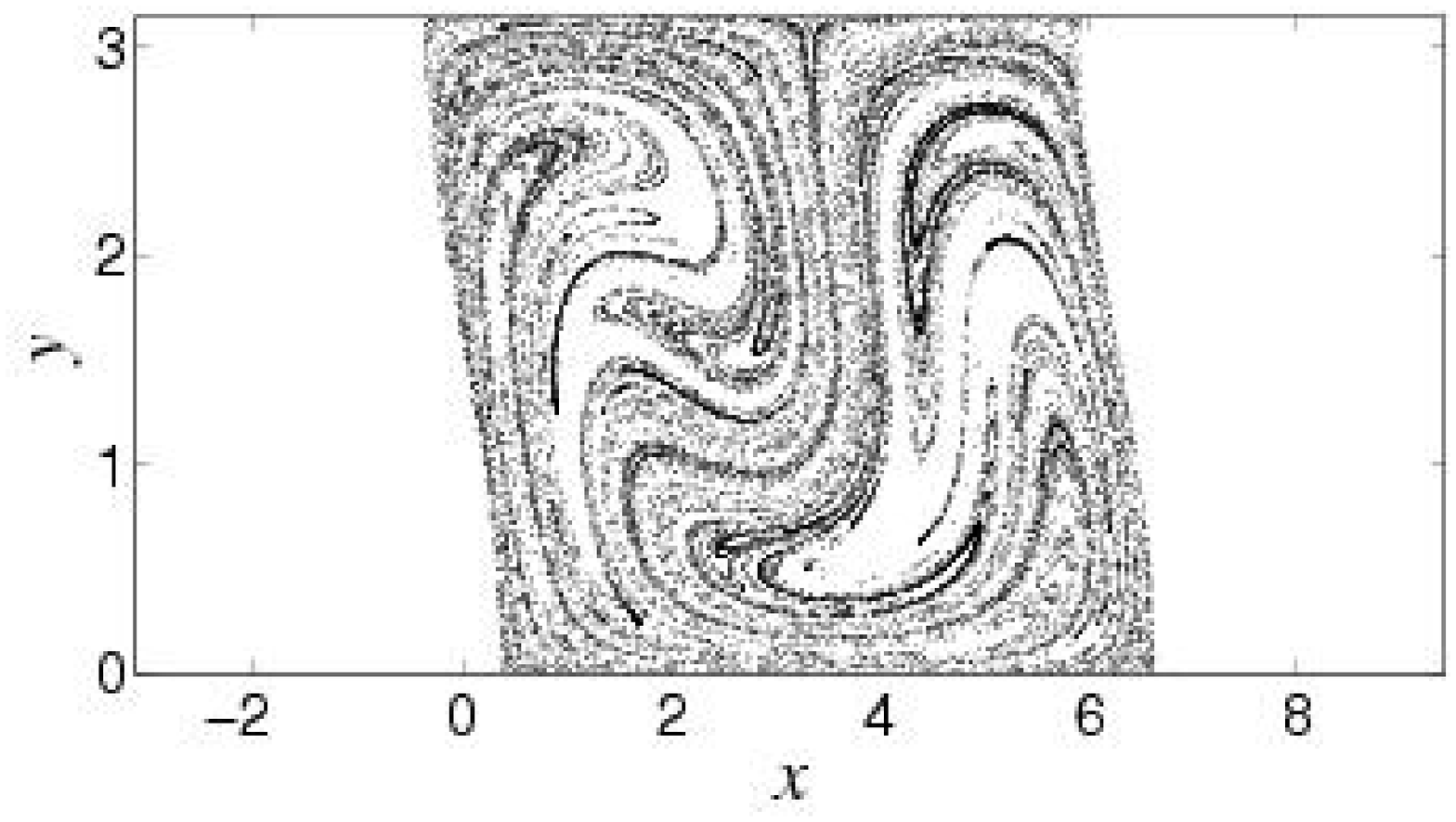}

\includegraphics[%
  width=3cm]{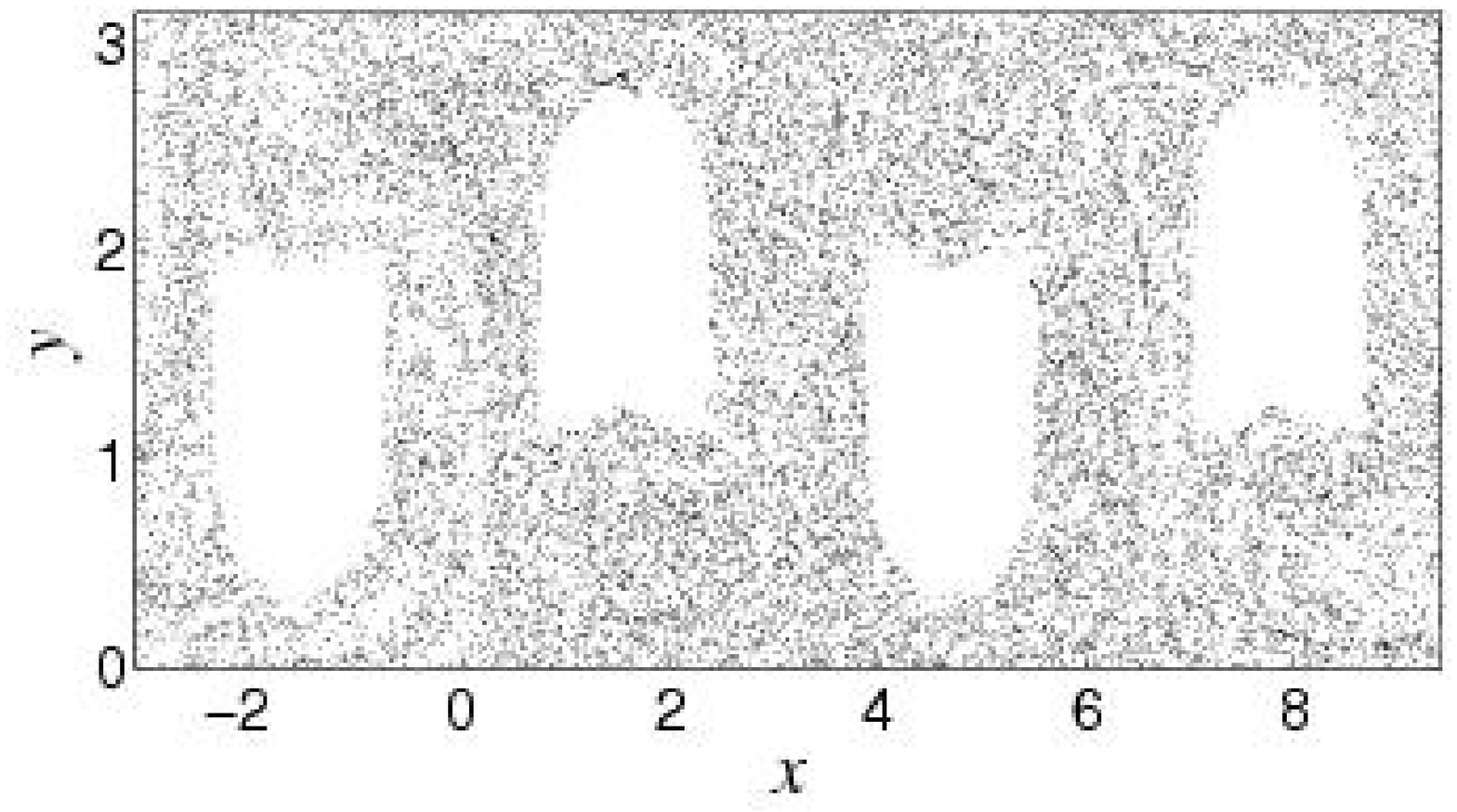}\includegraphics[%
  width=3cm]{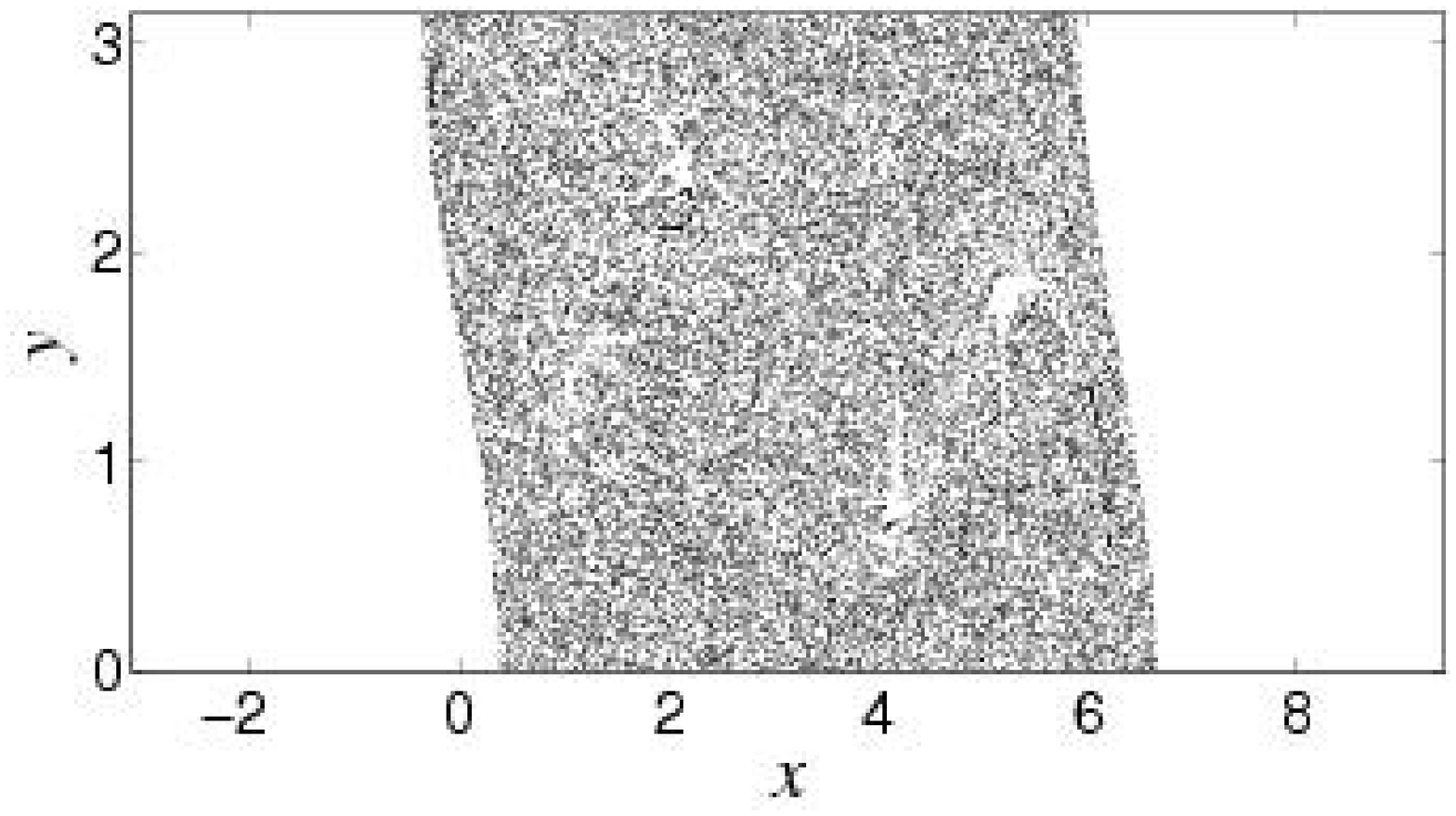}

\caption{\label{Fig4} Numerical simulation of the dynamics of a dye at $t=30$,
$t=50$, $t=70$ and $t=140$ (from top to bottom)~: left column
for the stream function~(\ref{courant}) and right column for the
stream function~(\ref{Strmcont}). The parameters are $\alpha=0.6$
and $\epsilon=0.63$.}
\end{figure}

In Fig.~\ref{Fig4}, we depict a numerical simulation of the dynamics of a dye in
the fluid. The left column shows the evolution of the tracers for
the stream function $\Psi_{1}$ given by Eq.~(\ref{courant}). The
right column shows the mixing of a dye within a cell which is limited by two
barriers created by the stream function $\Psi_{c}$ given by Eq.~(\ref{Strmcont}).
We see that the mixing occurs through a combination of stretching
and folding of the dye. The corresponding figure for $\Psi_{s}$ given by Eq.~(\ref{psis})
shows some advected particles outside the cell but the same picture inside.

In conclusion, we considered a situation of time-dependent oscillating
vortex chain. We have shown by
adding a suitable perturbation (forcing) it is possible to suppress the chaotic advection along the channel by creating dynamical barriers among the vortex chain, and at the same time, to enhance the mixing of passive tracers inside each of these isolated cells. The effect is obtained in spite of the oscillations of the rolls and is not too much affected by the use of a simplified version of the exact stream function. 

\begin{acknowledgments}
This work is supported by Euratom/CEA (contract EUR 344-88-1 FUA F).
We acknowledge useful discussions with M. Pettini, Y. Elskens, S.
Boatto, A. Goullet and the Nonlinear Dynamics group at CPT. 
\end{acknowledgments}

\end{document}